\documentclass[preprint,showpacs,preprintnumbers,amsmath,amssymb]{revtex4}
\usepackage{graphicx, epsfig}
\begin{document}


\title{Room Temperature Magnetocaloric Effect in Ni-Mn-In}
\author{P. A. Bhobe}
\address{Tata Institute of Fundamental Research, Homi Bhabha Road, Mumbai-400 005
 India.}
\author{K. R. Priolkar}
\address{Department of Physics, Goa University, Taliegao-Plateau, Goa-403 206 India.}
\author{A. K. Nigam}
 \address{Tata Institute of Fundamental Research, Homi Bhabha Road, Mumbai-400 005
 India.}
\date{\today}

\begin{abstract}
We have studied the effect of magnetic field on a non-stoichiometric
Heusler alloy Ni$_{50}$Mn$_{35}$In$_{15}$ that undergoes a
martensitic as well as a magnetic transition near room temperature.
Temperature dependent magnetization measurements demonstrate the
influence of magnetic field on the structural phase transition
temperature. From the study of magnetization as a function of
applied field, we show the occurrence of inverse-magnetocaloric
effect associated with this magneto-structural transition. The
magnetic entropy change attains a value as high as 25 J/kg-K (at 5 T
field) at room temperature as the alloy transforms from the
austenitic to martensitic phase with a concomitant magnetic
ordering.
\end{abstract}
\maketitle

In recent years, there has been an increasing awareness towards
incorporating an alternate technology for refrigeration that would
replace the conventional gas compression/expansion technique. In
this respect, the promising candidates are the magnetic materials
that exhibit an inherent {\it magnetocaloric effect} (MCE). With its
potential impact on energy saving and environment-friendly
applicability these magnetic materials are also looked up as
economically viable option for domestic and industrial
refrigeration.  Fundamentally, MCE is a process in which an
isothermal variation in the magnetic entropy takes place with an
adiabatic temperature change of the system on the application of
magnetic field. In this regard, many rare-earth and transition-metal
based alloys have found to show promising results. These include the
alloys like MnFeP$_{1-x}$As$_x$, Gd$_5$(SiGe)$_4$, MnAs$_{1-x}$Sb,
La(Fe$_x$Si$_{1-x}$)$_{13}$ \cite{teg,wada,pech,zhan,fuji}. The
research for the development of such MCE materials is made primarily
with two objectives: high MCE around room temperature, low-cost
production. Ferromagnetic Heusler alloy, Ni$_2$MnGa which pertains
to the class of shape memory alloy have been attracting much
attention in view of the giant MCE effect \cite{pareti}. The factor
responsible for the large entropy change is the first-order
structural and concomitant magnetic transition taking place in the
non-stiochiometric composition of this material, also giving rise to
the possibility of a magnetic control of shape memory effect. The
latest candidates in the field of such materials has been alloys
with composition Ni$_{50}$Mn$_{50-x}$Z$_x$ with Z = Sn, In, Sb
\cite{suto}. These materials share a number of features with
Ni-Mn-Ga alloys and also display interesting magnetic properties
\cite{acet1, acet2}. Being rich in Mn content (as compared to the
Heusler composition Ni$_2$MnZ), there are two crystallographic sites
occupied by Mn atoms giving rise to competing antiferromagnetic (AF)
exchange along with the underlying ferromagnetic (FM) order
\cite{brown18}. A certain composition of the Ni-Mn-Sn and Ni-Mn-In
alloys are shown to exhibit an {\it inverse} MCE effect in the
vicinity of martensitic to austenitic structural transition
\cite{acet-nature, khan}. However, the temperatures where such a
phenomenon is observed are much below the room temperature. In this
paper we present one such composition of Ni-Mn-In alloy that
undergoes a martensitic as well as a magnetic transition, both
occurring in a narrow temperature interval very close to the room
temperature. Owing to this fact, an appreciable change in the
magnetic entropy is observed at the transformation temperature. The
alloy has been thoroughly investigated for its magnetic properties
and the order of the sequential phase transitions have been clearly
determined.

Polycrystalline ingot of Ni-Mn-In was prepared by arc-melting the
starting elements (99.99\% purity) under argon atmosphere. To attain
a good compositional homogeneity, the ingots were re-melted 4-5
times with a weight loss of $\le$ 0.5\% and annealed at the
temperature of 900 K for 48 h in an evacuated quartz ampoule
followed by quenching in cold water. The precise composition of the
alloy was determined from energy dispersive x-ray (EDX) analysis to
be Ni = 50.4 at. \%; Mn = 34.5 at. \%; and In = 15.1 at. \%.
Magnetization (M) measurements were performed on a superconducting
quantum interference device magnetometer (Quantum Design, MPMS-5S)
in the temperature range 5 K to 330 K and magnetic field H = 0.01T
and 5T. M(H) up to 5T was measured at various temperatures near the
region of both, the martensitic and the magnetic transformations
with an interval of 2 K between each temperature value.

Magnetization as a function of temperature measured in low magnetic
field of 0.01 T, is presented in Fig. \ref{mag}(a). The sample was
initially cooled in the absence of field and data was collected on
warming from 5 to 330 K (ZFC), followed by cooling (FC) back to 5 K
and again during warming in the presence of field (FW). The
ferromagnetic transition is marked by the sharp rise in
magnetization at T$_C$ = 305 K. Below the magnetic ordering
temperature a sudden drop in magnetization takes place at 302 K with
value almost close to zero. This signature marks the formation of
the new structural phase. The drop in magnetization can be
conjectured to be due to the variants of the new crystallographic
phase being formed that temporarily disturbs the local ferromagnetic
orientation. This argument is further supported by the large
hysteresis in the temperature range of 265-310 K observed in ZFC, FC
and FW magnetization curves. Though the difference in ZFC and FC
curves can also be interpreted to be due to the nonzero coercivity
of the ferromagnetic material, the hysteresis between FC and
subsequent FW curves strongly suggests the presence of a first-order
structural transition within this temperature interval. Thus the
initial rise and the subsequent fall in magnetization with
decreasing temperature are signatures of the magnetic (T$_C$ = 305
K) and concomitant martensitic (T$_M \approx$  302 K) phase
transformations that the present Ni-Mn-In system undergoes. Below
150K, an irreversible behaviour between the ZFC and FC curves is
observed with ZFC response lying below the FC. Such irreversibility
is an indication of some degree of frustrated spin alignment. The
frustration of spins may result from the competition between the
underlying ferromagnetic coupling and an incipient antiferromagnetic
coupling between Mn atoms present at two different crystallographic
positions \cite{brown18}. The subsequent FW curve is seen to retrace
the FC curve over this temperature range.

Another striking feature in the magnetization curves is the decrease
in intensity of the peak-like signature in the vicinity of T$_M$ -
T$_C$ for the FW curve. This curve also shows a considerable
splitting in the narrow temperature interval. Such a signature
dictates the influence of magnetic field on the martensitic
transformation. Prior to the FW measurement, the sample undergoes a
cooling in some magnetic field (0.01 T in the present case) that
causes the T$_M$ to shift to a lower temperature value (295 K)
whereas the T$_C$ remains unchanged. This gives rise to the observed
splitting of the peak-like feature associated with martensitic and
magnetic transition in the M(T) curve. Fig. \ref{mag}(b) that
represents the FW M(T) measurement in magnetic field of 5 T, further
supports this observation. It is seen that the splitting of peak
observed at 295K in the FW curve, shifts to a lower value of 285 K
in the magnetic field of 5 T. Thus an overall shift in T$_M$ of
about -17 K is observed from the initial value of 302 K upon the
application of 5 T magnetic field. The lowering of T$_M$ in Ni-Mn-In
system implies that the magnetic field favors the formation of the
austenitic phase.

\begin{figure}
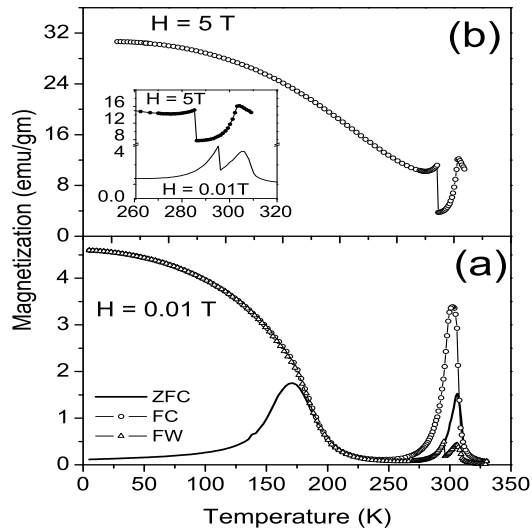

\caption{\label{arrott} Arrott Plots indicating the second order
magnetic and first order structural transition in
Ni$_{50}$Mn$_{35}$In$_{15}$.}
\end{figure}

To further investigate the magnetic properties of this alloy,
magnetization isotherms were measured for different temperatures
with an interval of 2 K, both above and below the two transition
temperatures. The measurements were carried out by cooling the
sample from 330 K down to the required temperature of interest and
then varying the field from 0 - 5 T. At the second order magnetic
transition, the magnetization isotherms presented in the Fig.
\ref{MH} show Brillouin-like dependence, with a slope that decreases
monotonically with increasing field. While the M(H) in the
temperature interval between 302 K and 296 K show prominent
meta-magnetic like characteristics that implies the first order
austenite to martensite structural phase transition. The M(H)
recorded at very low temperatures show a typical ferromagnetic
nature. In order to further assess the occurrence of first order
martensitic and second order magnetic transition in this sample, the
Arrott plots were plotted in the temperature range of 280 K$\le$ 320
K as shown in Fig. \ref{arrott}. For temperatures around T$_C$ = 305
K; the slopes are positive throughout, confirming the second-order
continuous nature of the transition in this sample. As the T$_M$ is
approached, the corresponding Arrott plots exhibit negative slopes.
This criterion has been perviously employed to determine the order
of sequential transitions in Ni-Mn-Ga alloys \cite{zhao} and
confirms the first-order nature of the martensitic transition at 302
K in the present Ni-Mn-In sample.

Finally, the magnetic entropy change $\Delta$S$_M$ has been
evaluated from the magnetization isotherms using the Maxwell
equation:
\begin{equation}
\Delta S_M(T,\Delta H) = \int_0^H\left(\frac{\partial
M(T,H)}{\partial T}\right)_H dH
\end{equation}

 The values of the magnetic entropy changes corresponding
to the average temperature between two consecutive M(H) isotherms,
calculated by numerical integration using the above expression over
the field span of 1 - 5 T are shown in Fig. \ref{mce}. A sharp peak,
with a $\Delta$S$_M$ value of 25 J/Kg-K is found at the temperature
of $\sim$ 301 K for the magnetic field variation from 1 to 5 T. The
obtained value of the maximum in $\Delta$S$_M$ is an excellent
feature particularly because it has been observed at room
temperature. This makes the Ni$_{50}$Mn$_{35}$In$_{15}$ alloy a
potential candidate for practical applications. This property
emerges as a consequence of spontaneous magnetization at room
temperature followed by the clear martensitic transition in the
narrow temperature interval.

In conclusion, we have observed a large magnetic entropy change
taking place at room temperature in Ni$_{50}$Mn$_{35}$In$_{15}$
alloy. A complete characterization of the magnetic properties of
this important material aids to the understanding required for the
technological exploitation of such materials. Especially, the narrow
difference ($\sim$ 2 K) between the structural and magnetic
transition temperature and alteration of T$_M$ by magnetic field are
remarkable properties that the present alloy exhibits.

\newpage
\begin{figure}
\caption{\label{mag} Thermal variation in magnetization of
Ni$_{50}$Mn$_{35}$In$_{15}$ measured in an applied field of (a) 0.01
T. (b) 5 T. Hysteresis between the FC, FH and sudden rise in ZFC,
FC, FH near 300 K are signatures of structural and magnetic
transitions respectively. The influence of magnetic field on the
structural transition is evident through the shift observed in T$_M$
(shown in the inset).}
\end{figure}

\begin{figure}
\caption{\label{MH} Magnetization isotherms measured above and below
the magnetic and structural phase transition in
Ni$_{50}$Mn$_{35}$In$_{15}$. Some of the low temperature isotherms
are shown separately in the upper panel of the figure.}
\end{figure}

\begin{figure}
\caption{\label{mce} Magnetic entropy change associated with the
structural and magnetic transition for Ni$_{50}$Mn$_{35}$In$_{15}$.
A large value of 25 J/kg-K is observed at 301 K for 5 T magnetic
field.}
\end{figure}

\newpage
\begin{figure}
\centering \epsfig{file=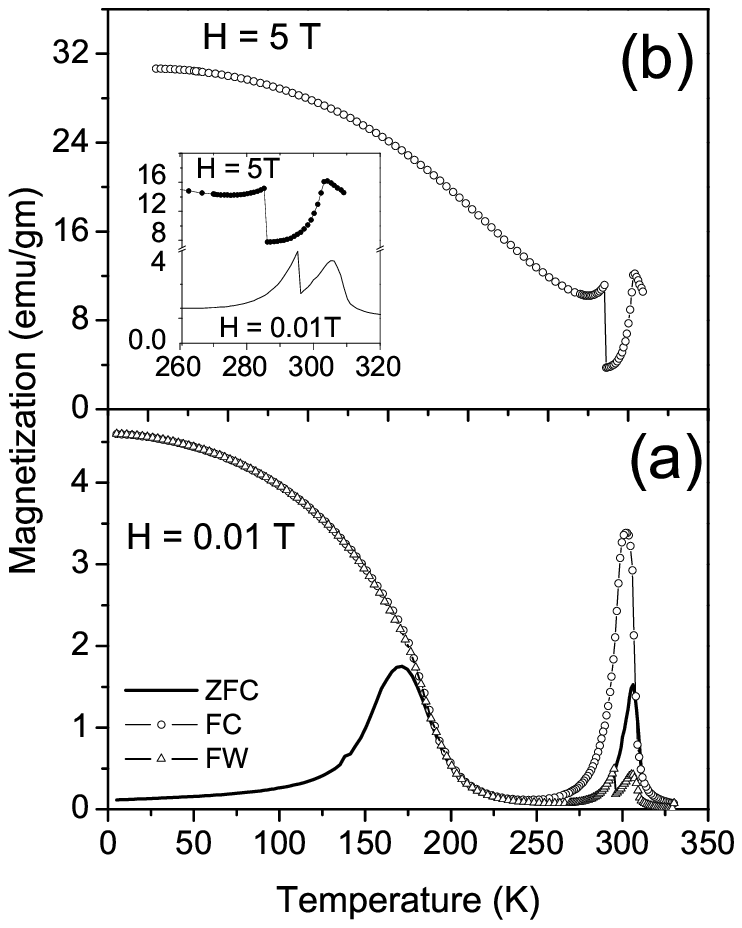, width=8cm, height=8cm}
\end{figure}

\begin{figure}
\centering \epsfig{file=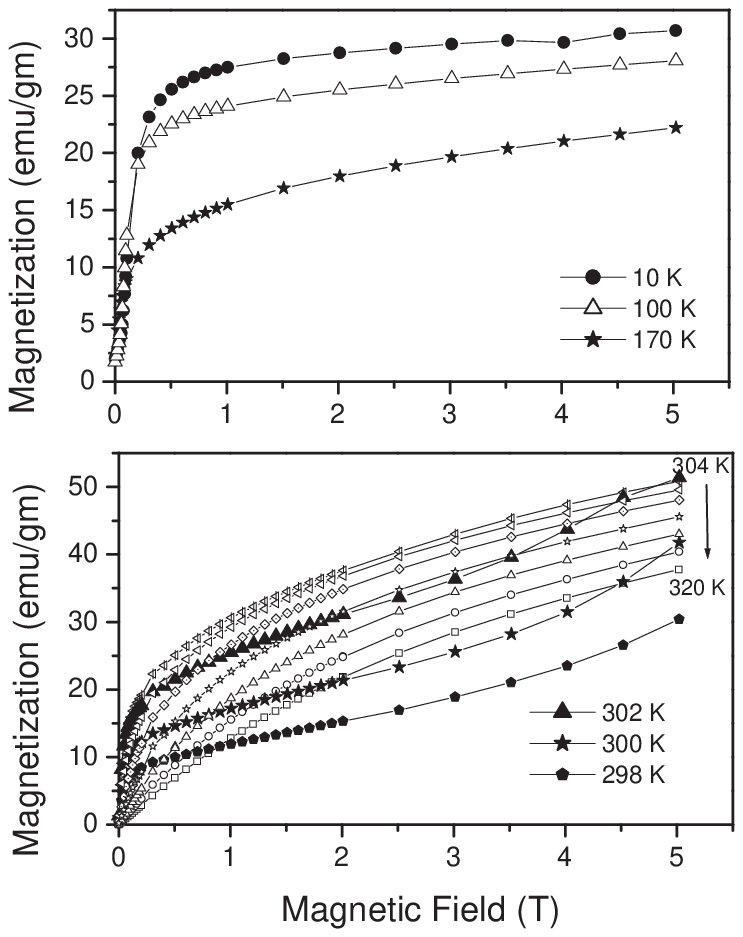, width=8cm, height=8cm}
\end{figure}

\begin{figure}
\centering \epsfig{file=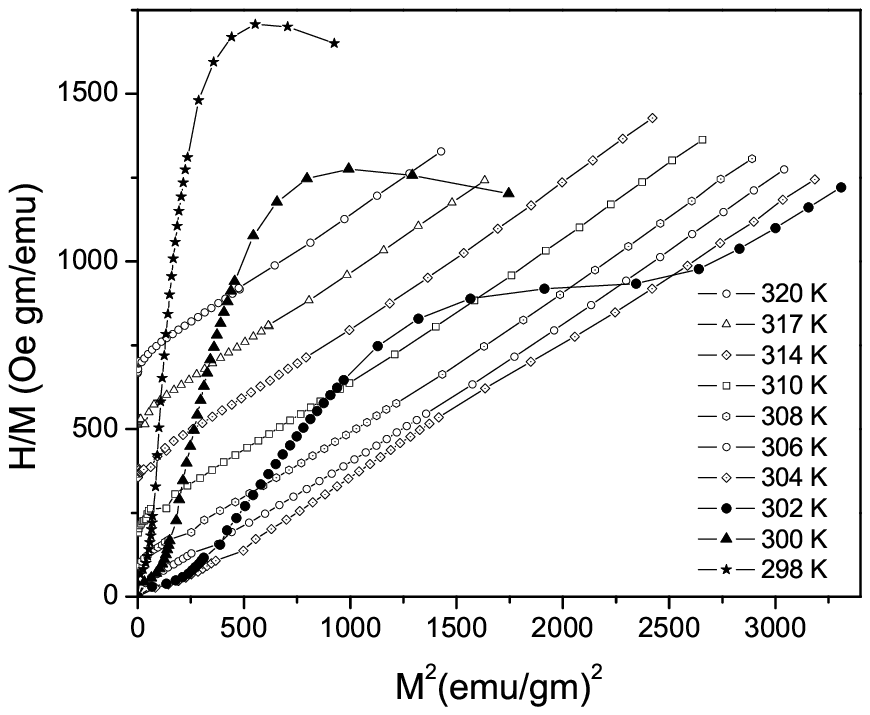, width=8cm, height=8cm}
\end{figure}

\begin{figure}
\centering \epsfig{file=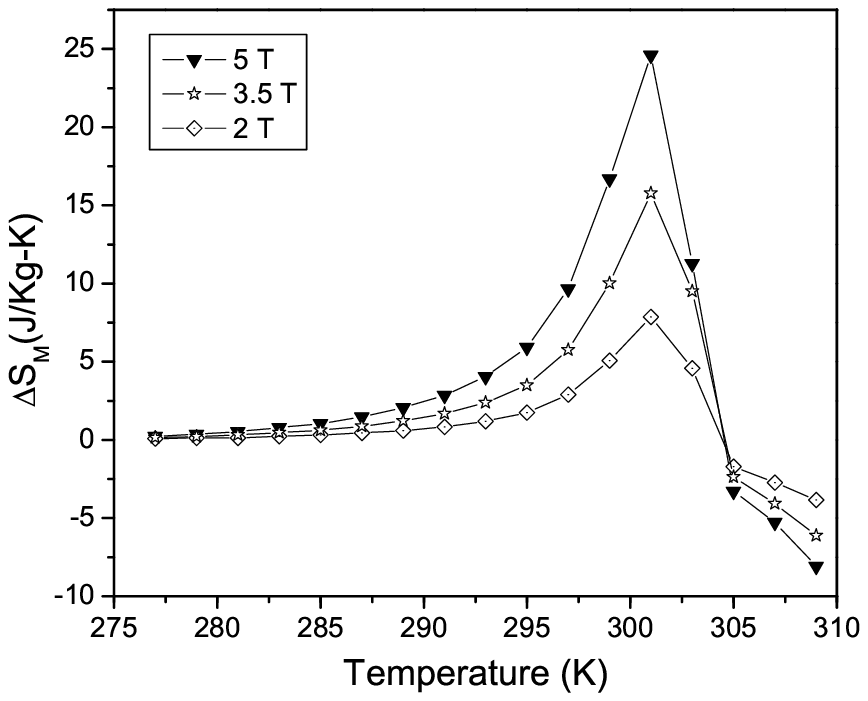, width=8cm, height=9cm}
\end{figure}


\newpage
\begin{thebibliography}{100}
\bibitem {teg} O. Tegus, E. Br\"{u}ck, K. H. Buschow and F. R. de
Boer, Nature {\bf 415}, 150 (2002).
\bibitem{wada} H. Wada, Y. Tanabe, Appl. Phys. Lett. {\bf 79} 3302
(2001).
\bibitem {pech} V. K. Pecharsky and K. A. Gschneidner, Jr., Phys.
Rev. Lett. {\bf 78}, 4494 (1997).
\bibitem{zhan} X. X. Zhang, G. H. Wen, F. W. Wang, W. H. Wang, C. H.
Yu, G. H. Wu, Appl. Phys. Lett. {\bf 77} 3072 (2000).
\bibitem{fuji} S. Fujieda, A. Fujita, K. Fukamichi, Appl. Phys.
Lett. {\bf 81} 1276 (2002).
\bibitem {pareti} L. Pareti, M. Solzi, F. Albertini and A. Paoluzi,
Eur. Phys. J. B. {\bf 32}, 303 (2003).
\bibitem {suto} Y. Sutou, Y. Imano, N. Koeda, T. Omori, R. Kainuma,
K. Ishida and K. Oikawa, Appl. Phys. Lett. {\bf 85}, 4358 (2004).
\bibitem {acet1} T. Krenke, M. Acet, E. F. Wassermann, X. Moya, L. Manosa and A. Planes,
Phys. Rev. B, {\bf 72}, 014412 (2005).
\bibitem {acet2} T. Krenke, M. Acet, E. F. Wassermann, X. Moya, L. Manosa and A. Planes,
Phys. Rev. B, {\bf 73}, 174413 (2006).
\bibitem {brown18} P. J. Brown, A. P. Grandy, K. Ishida, R. Kainuma, T. Kanomata,
K-U Neumann, K. Oikawa, B. Ouladdiaf and K. R. A. Ziebeck, J.
Phys.:Condens. Mater., {\bf 18} 2249 (2005).
\bibitem {acet-nature} T. Krenke, E. Duman, M. Acet, E. Wassermann, X. Moya, L. Manosa
and A. Planes, Nat. Mater. {\bf 4}, 450 (2005).
\bibitem{khan} A. K. Pathak, M. Khan, I. Dubenko, S. Stadler and N.
Ali, Appl. Phys. Lett. {\bf 90}, 262504 (2007).
\bibitem{zhao} X. Zhou, W. Li, H. P. Kunkel and G. Williams, Phys.
Rev. B, {\bf 73}, 012412 (2006).
\end{thebibliography}
\end{document}